\documentclass[aps,pra,twocolumn,nofootinbib]{revtex4-1}
                
\usepackage{bm,graphicx}
\usepackage{dsfont}
\usepackage{hyperref}
\usepackage[anythingbreaks]{breakurl}
\usepackage{amssymb,amsmath,amsfonts}
\usepackage{mathrsfs}
\usepackage{hhline}
\usepackage{epsfig}
\usepackage{epstopdf}
\usepackage[none]{hyphenat}

\usepackage[cp1250]{inputenc}
\DeclareInputText{175}{\.Z}     

\usepackage[mathscr]{eucal}
\usepackage{color}
\usepackage{marvosym}  

\usepackage[caption=false]{subfig}

\usepackage[cp1250]{inputenc}
\DeclareInputText{175}{\.Z}     

\usepackage[normalem]{ulem}

\newcommand*\xbar[1]{%
   \hbox{%
     \vbox{%
       \hrule height 0.5pt 
       \kern0.5ex
       \hbox{%
         \kern-0.1em
         \ensuremath{#1}%
         \kern-0.1em
       }%
     }%
   }%
} 

\newcommand{\id}{\mathds{1}}

\DeclareMathOperator{\Tr}{Tr}

\newcommand{\ket}[1]{\left|{#1}\right\rangle}
\newcommand{\bra}[1]{\left\langle{#1}\right|}

\newcommand{\rme}{\ensuremath{\mathrm{e}}}
\newcommand{\rmi}{\ensuremath{\mathrm{i}}}

\newcommand{\gd}{\ensuremath{\mathsf{g}}}
\newcommand{\hd}{\ensuremath{\mathsf{h}}}
\newcommand{\xd}{\ensuremath{\mathsf{x}}}
\newcommand{\yd}{\ensuremath{\mathsf{y}}}
\newcommand{\zd}{\ensuremath{\mathsf{z}}}
\newcommand{\ee}{\ensuremath{\mathsf{e}}}
\newcommand{\XX}{\ensuremath{\mathsf{X}}}
\newcommand{\ZZ}{\ensuremath{\mathsf{Z}}}
\newcommand{\DD}{\ensuremath{\mathsf{D}}}

\newcommand{\BHd}{\ensuremath{\mathcal{B}(\mathcal{H}_d)}}
\newcommand{\swap}{\ensuremath{\mathsf{SWAP}}}

\newcommand{\matel}[2]{\left|#1\middle\rangle\middle\langle#2\right|}

\begin{document}
%
\title{
       On orthogonal bases in the Hilbert-Schmidt space of matrices
      }

\author{Jens Siewert$^{1,2,3}$  
       }
\affiliation{
$^{1}$ Departamento de Qu\'{i}mica F\'{i}sica, Universidad del Pa\'{i}s Vasco UPV/EHU, E-48080 Bilbao, Spain}
\affiliation{
$^{2}$ EHU Quantum Center, Universidad del Pa\'{i}s Vasco UPV/EHU, E-48080
               Bilbao, Spain
}
\affiliation{
$^{3}$ IKERBASQUE Basque Foundation for Science, E-48013 Bilbao, Spain
}

\date{\today}
\begin{abstract}
    Decomposition of (finite-dimensional) operators in terms
    of  orthogonal bases of matrices has been a standard method in quantum
    physics for decades. In recent years, it has become increasingly popular
    because of various methodologies applied in quantum information, 
    such as the 
    graph state formalism and the theory of quantum error correcting codes,
    but also due to the intensified research on the Bloch
    representation of quantum states.
    In this contribution we collect various interesting facts and
    identities that hold for 
    finite-dimensional orthogonal matrix bases.
\end{abstract}
\maketitle

\section{Introduction}
%
Expansions of finite-dimensional matrices in terms of an orthogonal
basis of matrices as a method is ubiquitous in quantum physics, although 
it is often not explicitly mentioned (e.g., Refs.~\cite{Falkoff1951,Fano1954,Schwinger1960,Gell-Mann1961,Keldysh1965,Belzig1999,Winkler2003,CastroNeto2009,Cotfas2011,Vourdas2013,Vourdas2017}). 
In the past few decades, with the upcoming field of quantum information,
there has been increasing attention to techniques based on matrix expansions,
for example, in measurement-based quantum computation~\cite{Briegel2001}, 
the formalism of quantum error correcting codes~\cite{Gottesman1997,Knill1996,Scott2004}, the graph state formalism~\cite{Hein2006}, and entanglement 
theory \cite{Werner2000,CCNR,Myrheim2006}. There is also a
renewed interest in the Bloch representation, i.e., a decomposition
of the density operators of (usually) multipartite quantum states in terms 
of matrix bases~\cite{Fano1957,Eberly1981,Fano1983,Schlienz1995,Jaeger2003,Aschauer2004,deVicente2007,BertlmannKrammer2008,Kaszlikowski2008,deVicenteHuber2011,Kloeckl2015,Tran2016,Wyderka2019,ES2020}.

With this in mind, it is surprising that there do not seem to exist  surveys
on orthogonal matrix bases 
in the physics literature introducing basic definitions and 
known facts. While many isolated results, such as the 
expansions of the SWAP operator
or the maximally entangled state, are often used, e.g., in the 
quantum information literature we were not 
able to find a comprehensive discussion of those results and  the links
between them. The present article is a first attempt to fill this gap.
We focus on collecting the most important identities, 
presenting simple derivations of the well-known and also some new equalities, 
and exhibiting the relations both between them as well as to several other
interesting results.

We start by introducing the most frequently used matrix bases and then discuss
the transformation between arbitrary bases. Subsequently we analyze the 
matrix expansions of the two most important operators in this context,
the SWAP operator and the projector onto the Bell state, which give rise
to identities containing sums of products of two basis elements. But there
exist also relations for products with a larger number of factors. Finally we 
consider the explicit matrix representations of various interesting
maps. We elucidate the simple relations between
all these identities and hope to render transparent their sometimes surprising
structure.

\section{Orthogonal bases of matrices}%

%
\subsection{Definitions and frequently applied matrix bases}
\label{sec:defs}
%
In the following we discuss bases of the space $\mathcal{B}(\mathcal{H}_d)$
of bounded linear operators acting on a  
$d$-dimensional Hilbert space $\mathcal{H}_d$.
The Hermitian
adjoint of an operator $A\in\mathcal{B}(\mathcal{H}_d)$ 
is denoted by $A^{\dagger}$.
The space $\mathcal{B}(\mathcal{H}_d)$ is endowed with the 
Hilbert-Schmidt inner product
\begin{align}
       (A,B)_{\mathrm{HS}}\ =\ \Tr\left(A^{\dagger}B \right)\ \ ,
\end{align} 
where $\Tr$ denotes the trace. 
Correspondingly, the norm induced by this inner product is called
the Hilbert-Schmidt norm (also known as the Frobenius norm).
An operator basis in $\BHd$ has $d^2$
elements, therefore it can be represented by a  $d\times d$ matrix.
We will use the terms `operator' and `matrix' 
interchangeably, as it is common in the context of finite-dimensional
quantum mechanics. 
For the elements of the vector space $\mathcal{H}_d$
and the expansion of matrices in a given basis we will use the
Dirac {\em bra-ket} notation. That is, vectors $\psi\in \mathcal{H}_d$
are written as {\em kets} $\ket{\psi}$, whereas the elements 
$\varphi^{\ast}$ of the
dual space $\left(\mathcal{H}_d\right)^{\ast}$ are denoted as {\em bras},
$\bra{\varphi}$. A rank-1 matrix $\psi\varphi^{\ast}$ is written as
the outer product $\ket{\psi}\!\bra{\varphi}$. The standard basis in
$\mathcal{H}_d$ is called the {\em computational basis}
and denoted by $\{\ket{j}\}$, $j=0\ldots (d-1)$.

To start with an example, an operator (or matrix) basis every 
physicist is familiar with 
is given by the
Pauli matrices together with the $2\times 2$ identity matrix, 
an orthogonal basis for $\BHd$ in the case $d=2$,
\begin{align}
 \sigma_0\ =\ & \left[\begin{array}{cc}
                  1 & 0 \\
                  0 & 1
                 \end{array}
          \right]\ \ ,\ \
 \sigma_1\  =\  \left[\begin{array}{cc}
                  0 & 1 \\
                  1 & 0
                 \end{array}
          \right]\ \ , 
\nonumber\\
 \sigma_2\ =\ & \left[\begin{array}{cc}
                  0 & -\rmi \\
                  \rmi & 0
                 \end{array}
          \right]\ \ ,\ \
 \sigma_3\ =\  \left[\begin{array}{cc}
                  1 & 0 \\
                  0 & -1
                 \end{array}
          \right]\ \
\label{eq:Pauli} , 
\end{align}
with the normalization
\begin{align}
    \Tr\left(\sigma_j\sigma_k\right)\ =\ 2\delta_{jk}\ \ .
\end{align}
A well-known application of this basis is the expansion of a
density matrix $\rho^{(2)}$ of a two-level system, the so-called
Bloch representation for a qubit~\cite{Fano1954},
\begin{align}
       \rho^{(2)}\ =\ \frac{1}{2}\ \sum_j\ r_j\, \sigma_j\ \ ,
\end{align}
with real coefficients $r_j$ and $r_0=1$. There is an obvious 
second possibility to expand $\rho^{(2)}$ that makes reference to
the basis of the Hilbert (vector) space it acts on, the computational
basis $\{\ket{0},\ket{1}\}$,
\begin{align}
 \rho^{(2)}\ =\  \left[\begin{array}{cc}
                  \rho_{00} & \rho_{01} \\
                  \rho_{10} & \rho_{11}
                 \end{array}\right]
          \ =\ \sum_{j,k=0}^1\ \rho_{jk}\ \ket{j}\!\bra{k}\ \ .
\end{align}
That is, also the elements
\begin{align}
 \hat{\ee}_{00}= & \ket{0}\!\bra{0}\ \ ,\ \
 \hat{\ee}_{01}=\ket{0}\!\bra{1}\ \ ,\ \
\nonumber\\
 \hat{\ee}_{10}= & \ket{1}\!\bra{0}\ \ ,\ \
 \hat{\ee}_{11}=\ket{1}\!\bra{1}\ \ ,
\label{eq:standard2}
\end{align}
form an orthogonal basis of $\mathcal{B}(\mathcal{H}_2)$, however,
with normalization $\Tr\left(\hat{\ee}_{jk}^{\dagger}\hat{\ee}_{lm}\right)
=\delta_{jl}\delta_{km}$. This is the so-called standard basis.

By comparing Eqs.~\eqref{eq:Pauli} and~\eqref{eq:standard2}
we note that often for the enumeration of a matrix basis a single index 
$j=0\ldots (d^2-1)$ is
used (for example, also for $\lambda_1\ldots\lambda_8$, the Gell-Mann matrices 
in $d=3$). On the other
hand, the standard basis indicates that two indices would be a more natural choice
(or equivalently a two-digit base-$d$ number).
Throughout this paper we will adhere to the two-index enumeration.

For the sake of completeness we include here the general definitions for three
frequently used matrix bases in $d$ dimensions: the standard matrix basis, 
the 
Gell-Mann basis, and the Weyl operator basis. Importantly, we will normalize
all the matrix bases to the dimension $d$. As the basis of the underlying
Hilbert (vector) space $\mathcal{H}_d$ we use the computational 
basis $\{\ket{0},\ket{1} \ldots \ket{d-1}\}$. In contrast to the Hilbert space 
$\mathcal{H}_d$ we refer to $\BHd$ as the Hilbert-Schmidt space.

{\em The standard matrix basis.} It is straightforward to generalize Eq.~\eqref{eq:standard2}
to obtain the standard basis in $d$ dimensions (yet with different normalization),
\begin{align}
       \ee_{jk}\ =\ \sqrt{d}\ \ket{j}\!\bra{k}\ \ \ ,\ \ j,k=0,1\ldots (d-1)\ \ .
\label{eq:standardd}
\end{align}

{\em The (generalized) Gell-Mann basis.} The only matrix 
of this basis with non-vanishing trace
is $\Lambda_{00}^{(d)}\equiv \id_d$. The other $(d^2-1)$ traceless elements 
are Hermitian and correspond to the generators of SU($d$). They are defined by
\begin{subequations}
\begin{align}
  \label{eq:GMX}
  \Lambda _{kl}^{(d)} \equiv \xd_{kl} &=
  \sqrt{\frac{d}{2}}(\matel{k}{l}+\matel{l}{k})\ \ , \\
  \label{eq:GMY}
  \Lambda_{lk}^{(d)} \equiv \yd_{kl} &=
  \sqrt{\frac{d}{2}}(-\mathrm i\matel{k}{l}+\mathrm i\matel{l}{k})\ \ , \\
  \label{eq:GMZ}
  \Lambda_{ll}^{(d)} \equiv \zd_{ll} &= \sqrt{\frac{d}{l(l+1)}}
  \left(-l\matel{l}{l} + \sum_{j=0}^{l-1}\matel{j}{j}\right) \ , 
\end{align}
\label{eq:GMGM}
\end{subequations}
with $k<l=1\ldots (d-1)$. The Pauli basis is a special case of 
the Gell-Mann basis for $d=2$. Note that the ordering as well as the
normalization in $d=3$ is different from the standard one used in
high-energy physics.

{\em The Weyl operator basis.} Another highly interesting basis was introduced
in the quantum mechanics context by Weyl~\cite{Weyl1927} and
discussed also by Schwinger~\cite{Schwinger1960}. 
It consists of unitary matrices which are generated
by the {\em clock operator} $\ZZ$ and the {\em shift operator} $\XX$,
\begin{align}
        \ZZ\ \ket{j}\ =\ \omega^j\ \ket{j}\ \ \ ,\ \ \ \ 
        \XX\ \ket{j}\ =\ \ket{j+1}\ \ ,
\end{align}
where $\omega=\rme^{\frac{2\pi\rmi}{d}}$ and the addition is taken $\mod d$.
The basis elements are defined as
\begin{align}
        \DD_{jk}\ =\ \ZZ^j\ \XX^k\ \omega^{-\frac{jk}{2}}\ \ ,\ \ \ j,k=0\ldots(d-1)\ \ .
\end{align}
These operators are traceless. 
Moreover, they have a group structure. Also here, the Pauli matrices arise as the 
special case $d=2$, that is, they are both Hermitian and unitary.

There is an interesting observation regarding these three bases. Each of them
can be divided in two subsets, one of $d$ diagonal  and the other of $d(d-1)$
matrices with vanishing main diagonal. 

Another noteworthy aspect is the difference between the standard matrix basis
on the one hand, and Gell-Mann basis as well as Weyl operators on the other:
While the latter contain---except the identity matrix---only traceless elements
the former has several elements with nonvanishing trace. This property singles out 
matrix bases of $(d^2-1)$ traceless elements 
(such as the Gell-Mann and the Weyl operator bases) 
for the use in the Bloch decomposition of density matrices,
in particular of multipartite states. The reason is that the expansion
into traceless operators generates a structure into so-called 
sectors and subsectors~\cite{Wyderka2019,ES2020} that clearly
specify which parts of the expansion describe the reduced state across 
any possible cut in the set of parties.

%
%
%
\subsection{Transformation between matrix bases}
\label{sec:trafo}
%
A natural question is: What is the rule according to which two matrix
bases transform into one another? The answer is straightforward.
Suppose we are given two orthogonal matrix bases $\{\gd_{jk}\}$ and 
$\{\hd_{lm}\}$, $j,k,l,m=0 \ldots (d-1)$, with 
\begin{align*}
    \Tr\left(\gd^{\dagger}_{jk}\gd_{j'k'}\right)\ & =\ d\ \delta_{jj'}\delta_{kk'}\ \ ,
\\
\Tr\left(\hd^{\dagger}_{lm}\hd_{l'm'}\right)\ & =\ d\ \delta_{ll'}\delta_{mm'}\ \ ,
\end{align*}
and a $d^2\times d^2$ matrix $S$,
\begin{align}
   \hd_{jk}\ =\ \sum_{lm}\ S_{jk,lm}\ \gd_{lm}\ \ .
\end{align}
Note that this relation does not describe an action of $S$ on the $\gd_{lm}$.
It simply is the coefficient matrix that determines the linear combinations 
of the $\gd_{lm}$ to produce
the new basis elements $\hd_{jk}$.
Application of the orthogonality relations yields
\begin{align}
   \delta_{jj'}\delta_{kk'}\ =\ \sum_{lm}\ S_{jk,lm}^*\ S_{j'k',lm}\  \ ,
\label{eq:unitary}
\end{align}
which is nothing but the definition of a unitary matrix acting on 
a two-party space $\mathbb{C}^d\otimes\mathbb{C}^d$. 
What is of particular interest for the following is the transformation
between an arbitrary basis $\{\gd_{jk}\}$ and the standard basis, 
\begin{subequations}
\begin{align}
   \gd_{jk}\ =\ & \sqrt{d}\ \sum_{lm}\ U_{jk,lm}\ \ket{l}\!\bra{m}\ \ ,
\\
   \sqrt{d}\ \ket{j}\!\bra{k}\ =\ &   \sum_{lm}\ U_{jk,lm}^*\ \gd_{lm}\ \ ,
\end{align}
\label{eq:Ustandard}
\end{subequations}
with a two-party unitary matrix $U$~\cite{Karol2017,Bengtsson2017} 
as described above. If $\{\gd_{jk}\}$ has
the property that there are $d$ diagonal elements and the others with
vanishing diagonal, the matrix $U$ has  block structure: There is a
$d\times d$ block $\mathcal{D}$ transforming between the diagonal elements
of $\{\gd_{jk}\}$ and the $\ket{l}\!\bra{l}$, $l=0\ldots (d-1)$,
and another $d(d-1)\times d(d-1)$ block $\mathcal{O}$ for the offdiagonal
elements.
%
%
%
%
%
%
%
%


\section{SWAP operator and maximally entangled state}
%
In this section we analyze various matrix decompositions
that are related to the SWAP operator. As the SWAP decomposition is one
of the most frequently occurring `tricks' it is the appropriate
example to highlight the fact that the origin of the structure of many matrix basis
expansions is the unitary transformation law, Eq.~\eqref{eq:Ustandard}.
%
%
\subsection{Decomposition of the SWAP operator}
\label{sec:SWAP}

%
%
The SWAP operator exchanges the parties of a pure two-party
product state. Let $\ket{\psi},\ket{\varphi}\in \mathcal{H}_d$, then
\begin{align}
   \swap\ \ket{\psi}\otimes\ket{\varphi}\ =\ \ket{\varphi}\otimes\ket{\psi}\ \ .
\end{align}
The starting point is the decomposition of SWAP in the computational basis,
\begin{align}
   \swap\ =\ \sum_{jk}\ \ket{jk}\!\bra{kj} \ = \ \sum_{jk} \ket{j}\!\bra{k}\otimes
                                                           \ket{k}\!\bra{j}\ \ .
\label{eq:SWAPstandard}
\end{align}
By applying Eqs.~\eqref{eq:unitary}, \eqref{eq:Ustandard} 
we find the decomposition in another matrix basis $\{\gd_{lm}\}$,%
%
%
\footnote{In a high-energy physics context (e.g., Ref.~\cite{Greiner1994}) 
this relation is often called the
completeness relation for the generators of SU($d$) and spelled out in components
(the dagger could be dropped because of the hermiticity of the generators)
\[       \frac{1}{d}\
         \sum_{jk}\ \left(\gd_{jk}\right)_{lm} (\gd_{jk}^{\dagger})_{pq}\ =\
                    \delta_{lq}\delta_{mp}\ \ .
\]
}
\begin{align}
   \swap\ =\ & \frac{1}{d}\ \sum_{jklmpq}
          U^*_{jk,lm}\gd_{lm}\otimes \left(U^*_{jk,pq}\gd_{pq}\right)^{\dagger}
\nonumber\\
          =\ & \frac{1}{d}\ \sum_{lm}\ \gd_{lm}\otimes\gd_{lm}^{\dagger}\ \ .
\label{eq:SWAP}
\end{align}
The dagger in Eq.~\eqref{eq:SWAP} could equally well be assigned to the first 
tensor factor. The simplicity of the `diagonal' decomposition in any orthogonal
matrix basis hinges upon the unitarity of the transformation matrix $U$. It is the
same reason why the maximally entangled state 
has the same form for any local basis (and therefore
is called `isotropic'~\cite{Horodecki1999}). 
While the decomposition in Eq.~\eqref{eq:SWAP} might
appear surprising, the unitarity of $U$ implies that it can be read as if it were
written in the standard matrix basis -- where it is not surprising at all.

We may comment also on the case that the unitary $U$ has the block structure 
discussed in Sec.~\ref{sec:trafo} and the basis $\{\gd_{jk}\}$ can be subdivided in diagonal
and strictly offdiagonal matrices. By appropriately choosing the order of the columns in $U$
we can achieve that the matrices $\gd_{jj}$ are diagonal and
\begin{subequations}
\begin{align}
    \gd_{jj}\ & =\  \sqrt{d}\ \sum_{k}\ \mathcal{D}_{jk} \ket{k}\!\bra{k}
\\
  \sqrt{d}\  \ket{j}\!\bra{j}\ & =\  \sum_{k}\ \mathcal{D}_{jk}^*\ \gd_{kk}
\end{align}
\end{subequations}
with the unitary $d\times d$ matrix $\mathcal{D}$. From this we can infer that
for the diagonal part (in the matrix representation) of SWAP alone there is a  
matrix decomposition analogous to Eq.~\eqref{eq:SWAP},
\begin{align}
    \swap_{\mathrm{diag}}\ =\ & \sum_j\ \ket{jj}\!\bra{jj}
\nonumber\\
                           =\ & \frac{1}{d}\ \sum_{k}\ \gd_{kk}\otimes\gd_{kk}^{\dagger}
\ \ . 
\end{align}
An analogous expression can be found for the offdiagonal part of SWAP.

The two-party property of SWAP gives rise to another interesting relation
if we apply the partial trace, e.g., on the first party on the left-hand side
of Eq.~\eqref{eq:SWAP},
\begin{align}
  d\  \Tr_{[1]}\swap\ =\ 
            \sum_{lm}\ \Tr\left(\gd_{lm}\right)\ \gd_{lm}^{\dagger}\ =\ d\ \id_d\ \ ,
\label{eq:reducSWAP}
\end{align}
which implies also
\begin{align}            
\sum_{lm}\ \Tr\left(\gd_{lm}\right)\ \gd_{lm}^*\ =\ 
             \sum_{lm}\ \Tr\left(\gd^*_{lm}\right)\ \gd_{lm}^T\ =\ d\ \id_d\ \ ,
\label{eq:TrCoeff}
\end{align}
where $T$ denotes the transposition.  Tracing out the remaining party leads to
\begin{align}
   d^2\ =\ \sum_{lm}\ \left|\Tr\left(\gd_{lm}\right)\right|^2\ \ .
\label{eq:Tr12Bell}
\end{align}

Finally, by looking at Eq.~\eqref{eq:SWAPstandard} one might wonder whether
it is possible to get an analogous matrix expansion also for the identity operator
$\id_d\otimes\id_d=\sum_{jk} \ket{jk}\!\bra{jk}$. The answer is affirmative and
can be found by exploiting the relation
\begin{align}
    \swap\ \cdot\ \swap\ =\ \id_d\otimes\id_d\ \ .
\label{eq:SWAPSWAP}
\end{align}
Substituting Eq.~\eqref{eq:SWAP} gives
\begin{align}
    \id_d\otimes\id_d\ =\ 
           \frac{1}{d^2}\ \sum_{jkab}\ \gd_{ab}^{\dagger}\cdot\gd_{jk}
                                           \otimes
                                       \gd_{ab}\cdot\gd_{jk}^{\dagger}\ \ ,
\label{eq:identity}
\end{align}
where again several combinations of placing the daggers are possible. The structure
of this equation is somewhat unexpected, because in general a matrix basis does
not have group properties, that is, the products $\gd_{ab}^{\dagger}\cdot\gd_{jk}$
bear no obvious relation with one another and in particular 
do not reproduce the elements of a basis (consider the case, e.g., that the $\gd_{jk}$
are the Gell-Mann matrices).
Also here we can take the partial trace as in Eq.~\eqref{eq:reducSWAP} and obtain
\begin{align*}
    d^3\ \id_d\ =\ 
            \sum_{jkab}\ \Tr\left(\gd_{ab}^{\dagger}\cdot\gd_{jk}\right)
                                       \gd_{ab}\cdot\gd_{jk}^{\dagger}\ \ ,
\end{align*}
so that 
\begin{align}
    d^2\ \id_d\ =\ 
            \sum_{jk}\ 
                                       \gd_{jk}\cdot\gd_{jk}^{\dagger}\ \ .
\label{eq:Tr1-identity}
\end{align}
%
%
%
%
%
%
\subsection{Further results related to the SWAP operator}
\label{sec:furtherSWAP}
%
There is an intriguing identity
including the SWAP operator~\cite{Jamiolkowski1972}, which is
at the heart of the Choi-Jamiolkowski isomorphism, namely
\begin{align}
    \Tr_{[2]}\left(A\otimes B\cdot \swap\right)\ =\ A\cdot B\ \ , 
\label{eq:TrSWAP}
\end{align}
with single-party operators $A,B\in \BHd$. 
It is readily verified by 
expanding SWAP in the standard matrix basis. This is also helpful
to elucidate the workings of this relation:
$ \Tr_{[2]}\left(A\otimes B\cdot \swap\right)=\sum_{jk} A\ket{j}\!\bra{k}\bra{j}B\ket{k}
=\sum_{jk} A\ket{j}\!\bra{j}B\ket{k}\!\bra{k}$.

Alternatively, we may look at Eq.~\eqref{eq:TrSWAP} from the point of view of the
decomposition Eq.~\eqref{eq:SWAP}. To this end, it is useful to keep in mind the
Bloch representation of $B$ in the matrix basis $\{\gd_{jk}\}$,
\begin{align}
      B\ =\ \frac{1}{d}\ \sum_{jk}\ b_{jk}\gd_{jk}\ \ \  ,\ \ \ 
      b_{jk}\ =\ \Tr\left(\gd_{jk}^{\dagger}B\right)\ .
\label{eq:Blochdec}
\end{align}
With this we find 
\begin{align}
    \Tr_{[2]}\left(\id\otimes B\cdot \swap\right)\ & =\ 
                                 \frac{1}{d} \sum_{jk} \gd_{jk}
                                                       \Tr\left(B\ \gd_{jk}^{\dagger}\right)
\nonumber\\
                                                   & =\ 
                                                        B\ \ ,
\label{eq:TrSWAPId}
\end{align}
so that Eq.~\eqref{eq:TrSWAPId} can be interpreted in the sense of decomposing $B$ into its
Bloch components on the second party and rebuilding it on the first (and Eq.~\eqref{eq:TrSWAP}
follows trivially through multiplication by $A$ from the left). If we set $A=B^{\dagger}$ 
and take
the trace over both parties separately in Eq.~\eqref{eq:TrSWAP} we immediately
find with the matrix decomposition of SWAP
\begin{align}
    \Tr\left(B^{\dagger}\otimes B\cdot \swap\right) = \frac{1}{d} \sum_{jk} |b_{jk}|^2
                                           = \Tr\left(B^{\dagger} B\right)\  ,
\end{align}
which links the length of the Bloch vector of $B$ to the purity of the operator.

Proceeding along these lines we can combine Eqs.~\eqref{eq:SWAP}, 
\eqref{eq:SWAPSWAP}
and~\eqref{eq:TrSWAP},
\begin{align}
    d^2\ \id_d\ & =\ d\ \Tr_{[2]}\left(\swap \cdot \swap\right)
\nonumber\\
              & =\ \sum_{jk}\Tr_{[2]}\left(\gd_{jk}\otimes\gd_{jk}^{\dagger}
                                         \cdot \swap\right)
\nonumber\\
              & =\ \sum_{jk}\ \gd_{jk}\cdot\gd_{jk}^{\dagger}
\nonumber\\
              & =\ \sum_{jk}\ \gd_{jk}^{\dagger}\cdot\gd_{jk}\ \ ,
\label{eq:skewSWAP}
\end{align}
which again leads us to Eq.~\eqref{eq:Tr1-identity}, with a slightly different
interpretation.
This equality demonstrates that the elements of a matrix basis 
constitute (up to prefactors) the Kraus
operators of a unital channel, the depolarizing channel.

By reading Eq.~\eqref{eq:skewSWAP} in terms of the standard basis we see
that there is a more direct way to obtain this identity: We could have
used $d^2\id_d=d\sum_{jk} 
\ket{j}\!\bra{k}\cdot\left( \ket{j}\!\bra{k}\right)^{\dagger}$ and 
applied the transformation Eq.~\eqref{eq:Ustandard}.

Moreover, it is also possible to derive equalities that contain inner products
of four basis elements.
For this purpose, we substitute Eq.~\eqref{eq:identity} into Eq.~\eqref{eq:SWAP} and find
\begin{align}
    \id_d\ & =\ \Tr_{[2]}\left(\id_d\otimes\id_d\cdot\swap\right)
\nonumber\\
         \ & =\ \frac{1}{d^2}\ \sum_{jkab}\ \Tr_{[2]}\left(\gd_{ab}^{\dagger}\cdot\gd_{jk}
                                           \otimes
                                       \gd_{ab}\cdot\gd_{jk}^{\dagger}\cdot\swap\right)
\nonumber\\
         \ & =\ \frac{1}{d^2}\ \sum_{jkab}\ \gd_{ab}^{\dagger}\cdot\gd_{jk}\cdot
                                       \gd_{ab}\cdot\gd_{jk}^{\dagger}\ \ .
\label{eq:fourops1}
\end{align}
%
%
%
%
%
%
\subsection{The maximally entangled state}
\label{sec:Bell}
%
The maximally entangled state (or Bell state) in a $d\times 
d$-dimensional Hilbert space is defined as
\begin{align}
 \ket{\Phi^+_d}\ =\ \frac{1}{\sqrt{d}}\ \sum_j\ \ket{jj}\  \ .
\label{eq:Bellstate}
\end{align}
From Eq.~\eqref{eq:SWAPstandard} we can immediately infer the relation to the SWAP operator
\begin{align}
  d  \ket{\Phi^+_d}\!\bra{\Phi^+_d}\ =\  \sum_{jk}\ \ket{jj}\!\bra{kk}\
                                   =\  \swap^{T_{[2]}}\ \ ,
\end{align}
where $T_{[2]}$ denotes the partial transposition on the second party of
the two-party SWAP. Therefore, the matrix decomposition of the Bell state in
the basis $\{\gd_{lm}\}$ takes the form
\begin{align}
   \ket{\Phi^+_d}\!\bra{\Phi^+_d}\ =\  
          \frac{1}{d^2}\ \sum_{jk}\ \gd_{jk}\otimes\gd_{jk}^*\ \ ,
\label{eq:Bell}
\end{align}
which again is diagonal in the matrix indices. Obviously the partial transposition
could be done as well on the first  party, so that the single term in the sum
would read $\gd_{lm}^T \otimes\gd_{lm}^{\dagger}$, and also the other combination
of transposition and Hermitian adjoint (or only complex conjugation) 
is possible.

We mention the interesting special case of Eq.~\eqref{eq:Bell} when the 
operator basis consists of two subsets of matrices, one with strictly real 
and the other with strictly imaginary coefficients.
This holds for the Gell-Mann basis, Eq.~\eqref{eq:GMGM}. 
Here, the complex conjugation 
in Eq.~\eqref{eq:Bell} only changes the sign of the terms of the
imaginary matrices, so that
\begin{align}
 \ket{\Phi^+_d}\!\bra{\Phi^+_d}\ =\ & \frac{1}{d^2}
           \sum_{j<k}\ \bigg[\xd_{jk}\otimes\xd_{jk}
                                       -   \yd_{jk}\otimes\yd_{jk} \bigg] +
\nonumber\\
          & \ \ \ \ \ \  +\  \frac{1}{d^2} \sum_{j} \zd_{jj}\otimes\zd_{jj}
\ \ ,
\label{eq:Bell-GM}
\end{align}
where $\zd_{00}\equiv\id_d$.

We can now derive additional matrix relations in analogy with Sec.~\ref{sec:SWAP}.
We have the identity
\begin{align*}
   \ket{\Phi^+_d}\!\bra{\Phi^+_d}\cdot  \ket{\Phi^+_d}\!\bra{\Phi^+_d}\ =\  
   \ket{\Phi^+_d}\!\bra{\Phi^+_d}\ \ ,
\end{align*}
and together with Eq.~\eqref{eq:Bell},
\begin{align}
   \ket{\Phi^+_d}\!\bra{\Phi^+_d} \ =\ 
           \frac{1}{d^4}\ \sum_{jkab}\ \gd_{ab}\cdot\gd_{jk}
                                           \otimes
                                       \left(\gd_{ab}\cdot\gd_{jk}\right)^*\ \ ,
\label{eq:BellBell}
\end{align}
where again there are several possibilities to place the complex conjugation.
Also this equation is remarkable, because it has the same structure as
Eq.~\eqref{eq:Bell}, while the products $\gd_{ab}\cdot\gd_{jk}$ do not necessarily form 
a basis. Taking the partial trace gives
\begin{align}
   d^3\ \id_d \ =\ 
            \sum_{jkab}\ \Tr\left(\gd_{ab}\cdot\gd_{jk}\right)\
                                       \left(\gd_{ab}\cdot\gd_{jk}\right)^*\ \ ,
\label{eq:Tr1BellBell}
\end{align}
while tracing out both parties leads to
\begin{align}
   d^4\ =\ \sum_{jkab}\ \left|\Tr\left(\gd_{ab}\cdot\gd_{jk}\right)\right|^2\ \ .
\label{eq:Tr12BellBell}
\end{align}
%
%
%
%
%
%
%
\subsection{Combining SWAP and  Bell state}
%
The Bell state is invariant under the permutation of its two parties, hence
\begin{align}
   \swap\cdot  \ket{\Phi^+_d}\!\bra{\Phi^+_d}\ & =\  
   \ket{\Phi^+_d}\!\bra{\Phi^+_d}\cdot\swap 
\nonumber\\
   \ & =  \
   \ket{\Phi^+_d}\!\bra{\Phi^+_d}\  .
\end{align}
First, this gives another relation analogous to Eq.~\eqref{eq:BellBell},
\begin{align}
   \ket{\Phi^+_d}\!\bra{\Phi^+_d} \ =\ 
           \frac{1}{d^3}\ \sum_{jkab}\ \gd_{ab}\cdot\gd_{jk}^*
                                           \otimes
                                  \gd_{ab}^{\dagger}\cdot\gd_{jk}
\ \ ,
\label{eq:SWAPBell}
\end{align}
From this, we obtain by tracing out the second party
\begin{align}
     d\  \id_d\ =\  \sum_{jk}\ \gd_{jk}\cdot\gd_{jk}^*\ \ .
\end{align}
%
%

By using Eq.~\eqref{eq:TrSWAP} we find two more equalities containing 
four-matrix products,
\begin{align}
   \id_d\ & =\  \frac{1}{d^3}\ \sum_{jkab}\ \Tr_{[2]}\bigg[\gd_{ab}\cdot\gd_{jk}
                                           \otimes
                           \left(\gd_{ab}\cdot\gd_{jk}\right)^*\cdot\swap\bigg]
\nonumber\\
          & =\ \frac{1}{d^3}\ \sum_{jkab}\ \gd_{ab}\cdot\gd_{jk}\cdot
                                       \gd_{ab}^*\cdot\gd_{jk}^*\ \ ,
\label{eq:fourops2}
\end{align}
and
\begin{align}
   \id_d\ & =\  \frac{1}{d^2}\ \sum_{jkab}\ \Tr_{[2]}\big[\gd_{ab}\cdot\gd_{jk}^*
                                           \otimes
                                 \left(\gd_{ab}^{\dagger}\cdot\gd_{jk}\right)\cdot\swap\big]
\nonumber\\
          & =\ \frac{1}{d^2}\ \sum_{jkab}\ \gd_{ab}\cdot\gd_{jk}^*\cdot
                                       \gd_{ab}^{\dagger}\cdot\gd_{jk}\ \ .
\label{eq:fourops3}
\end{align}
We will see below that these identities are closely related to those
in Eqs.~\eqref{eq:TrCoeff} and~\eqref{eq:skewSWAP}.
%

\subsection{The fully coherent state}
%
For the sake of completeness we mention here also a link 
between  matrix bases to the fully coherent state in $\mathcal{H}_d$,
\begin{align}
        \ket{+}_d\ =\ \frac{1}{\sqrt{d}}\ \sum_{j}\ \ket{j}\ \ .
\end{align}
In the standard basis we see that
\begin{align}
      d\  \ket{+}_d\!\bra{+}\ =\ \sum_{jk}\ \ket{j}\!\bra{k}\ \ ,
\end{align}
so that, by virtue of Eq.~\eqref{eq:Ustandard},
\begin{align}
      \sqrt{d}^3\  \ket{+}_d\!\bra{+}\ =\ \sum_{jklm}\ 
                          U^*_{lm,jk}\gd_{lm}\ \ .
\end{align}
The components of the unitary $U$ appear explicitly because
there is no double occurrence of matrix index pairs.

\subsection{Collecting the results}
%
Let us briefly summarize the main results
we have obtained so far. Starting from the decompositions
of SWAP and the Bell state, we found a number of relations that contain
products of two matrices,
\begin{align*}
 \swap\ =\ &  \frac{1}{d}\ \sum_{jk}\ \gd_{jk}\otimes\gd_{jk}^{\dagger}\ \ ,
\\
 d^2\ \id_d\ =\ &  \sum_{jk}\ \gd_{jk}\cdot\gd_{jk}^{\dagger}\ \ ,
\\
   d\ \id_d\ =\ &  \sum_{jk}\ \Tr\left(\gd_{jk}\right)\ \gd_{jk}^{\dagger}\ \ ,
\\
   d^2\ =\ &  \sum_{jk}\ \left|\Tr\left(\gd_{jk}\right)\right|^2\ \ ,
\end{align*}
and
\begin{align*}
   \ket{\Phi^+_d}\!\bra{\Phi^+_d} \ =\ 
      &  \frac{1}{d^2}\ \sum_{jk}\ \gd_{jk}\otimes\gd_{jk}^{\ast}\ \ ,
\\
 d\ \id_d\ =\ &  \sum_{jk}\ \gd_{jk}\cdot\gd_{jk}^{\ast}\ \ ,
\\
   d\ \id_d\ =\ &  \sum_{jk}\ \Tr\left(\gd_{jk}\right)\ \gd_{jk}^{\ast}\ \ ,
\\
   d^2\  =\ &  \sum_{jk}\ \Tr\left(\gd_{jk}\cdot\gd_{jk}^{\ast}\right)\ \ .
\end{align*}
Moreover, there are several identities with products of
four matrices
\begin{align*}
    \id_d\otimes\id_d\ =\ &
           \frac{1}{d^2}\ \sum_{jkab}\ \gd_{ab}^{\dagger}\cdot\gd_{jk}
                                           \otimes
                                       \gd_{ab}\cdot\gd_{jk}^{\dagger}\ \ ,
\\
    d^2\ \id_d\ =\ & \sum_{jkab}\ \gd_{ab}^{\dagger}\cdot\gd_{jk}\cdot
                                       \gd_{ab}\cdot\gd_{jk}^{\dagger}\ \ ,
\\
    d^3\ \id_d\ =\ & \sum_{jkab}\ \gd_{ab}\cdot\gd_{jk}\cdot
                                       \gd_{ab}^*\cdot\gd_{jk}^*\ \ ,
\\
   \ket{\Phi^+_d}\!\bra{\Phi^+_d} \ =\ &
           \frac{1}{d^3}\ \sum_{jkab}\ \gd_{ab}\cdot\gd_{jk}^*
                                           \otimes
                                       \gd_{ab}^{\dagger}\cdot\gd_{jk}\ \ ,
\\
                                  \ =\ &
           \frac{1}{d^4}\ \sum_{jkab}\ \gd_{ab}\cdot\gd_{jk}
                                           \otimes
                                       \left(\gd_{ab}\cdot\gd_{jk}\right)^*\ \ ,
\\
    d^2\ \id_d\ =\ & \sum_{jkab}\ \gd_{ab}\cdot\gd_{jk}^*\cdot
                                       \gd_{ab}^{\dagger}\cdot\gd_{jk}\ \ ,
\\
   d^4\ =\ & \sum_{jkab}\ \left|\Tr\left(\gd_{ab}\cdot\gd_{jk}\right)\right|^2\ \ .
\end{align*}
As discussed before, it is possible to re-assign conjugations, daggers and
transpositions to other factors in most of the relations.

\section{Decompositions of the trace and other maps}
\label{sec:maps}
%
Matrix expansions are relevant not only for operators like SWAP,
but also for quantum channels and other maps, as indicated already
in the preceding section. Here we provide some explicit examples
that are particularly useful in calculations using the Bloch
representation.
%
%
\subsection{The trace}
\label{sec:trace}
%
The trace maps an operator into the complex numbers, but instead one
may consider the map to another operator  $\rho \rightarrow \Tr(\rho)\ \id_d$.
There is a well-known matrix expansion for this map (cf., e.g., 
Ref.~\cite{Schwinger1960}). Consider an operator 
$A\in \BHd$. Then
\begin{align}
       \Tr(A)\ \id_d\ =\ \frac{1}{d}\ \sum_{lm}\ 
                               \gd_{lm}\cdot A\cdot \gd_{lm}^{\dagger}\ \ ,
\label{eq:trace}
\end{align}
which at first glance looks remarkable -- how can this work
for any matrix basis $\{\gd_{jk}\}$? Applying our method from the previous sections,
we start by reading this relation in the standard matrix basis.  We find that here
it is rather evident,
\begin{align*}
       \Tr(A)\ \id_d\ & =\ \sum_k\ \bra{k} A \ket{k}\otimes\id_d
\nonumber\\
                           & =\ \sum_{jk} \ket{j}\!\bra{k} A \ket{k}\!\bra{j}
\nonumber\\
                           & =\ \sum_{jk} \ket{j}\!\bra{k}\cdot A \cdot
                            \left(\ket{j}\!\bra{k}\right)^{\dagger}\ \ .
\end{align*}
As before, we notice the double occurrence of the basis indices, therefore we know
that substituting the transformation Eq.~\eqref{eq:Ustandard} straightforwardly
leads to the result,
\begin{align*}
       \Tr(A)\ \id_d\ & =\ \frac{1}{d}\ \sum_{jkmnpq} U^*_{jk,lm} \gd_{lm}
                                               \cdot A \cdot
                                       \gd_{pq}^{\dagger} U_{jk,pq}
\nonumber\\                & =\ \frac{1}{d}\ \sum_{lm} \ \gd_{lm}
                                               \cdot A \cdot
                                       \gd_{lm}^{\dagger} \ \ .
\end{align*}
After all, Eq.~\eqref{eq:trace} confirms what was to be expected from 
Eq.~\eqref{eq:skewSWAP} in Sec.~\ref{sec:furtherSWAP}: 
The depolarization channel can be realized
by using the elements of a matrix basis as Kraus operators.
With this expansion of the trace map it becomes obvious that Eq.~\eqref{eq:fourops3} 
can be read as a version of Eq.~\eqref{eq:TrCoeff} [the same holds for 
Eq.~\eqref{eq:fourops1}].

Evidently, in the multipartite case Eq.~\eqref{eq:trace} can be applied 
as partial trace operation to any subset of parties.
%
%
\subsection{The identity map}
\label{sec:identity}
In analogy with Sec.~\ref{sec:SWAP} a small detour has
to be taken in order to represent the identity map
in terms of a matrix basis, 
$\mathsf{Id}(A)=\id_d\ A\ \id_d=\sum_{jk} \ket{j}\!\bra{j}A\ket{k}\!\bra{k}$.
This is because the two indices do not belong to a single matrix basis.
We can achieve this by writing
\[ 
         \sum_{jklm}\ \ket{j}\!\bra{k}\cdot\ket{m}\!\bra{l}\cdot A\cdot 
                      \ket{k}\!\bra{j}\cdot\ket{l}\!\bra{m}\ \ ,
\]
%
which leads to
%
\begin{align}
       \mathsf{Id}(A)\ =\  & \frac{1}{d^2}\
                  \sum_{jklm}\ \gd_{jk}\cdot\gd_{lm}^{\dagger}\cdot A\cdot
                               \gd_{jk}^{\dagger}\cdot\gd_{lm}
\label{eq:id-dagger}
\ \ .
\end{align}
%
Equation~\eqref{eq:fourops1} 
may be viewed, for example, as the normalization condition for the map
in Eq.~\eqref{eq:id-dagger}.
In fact, we could have directly used Eq.~\eqref{eq:identity} and
the SWAP relation~\eqref{eq:TrSWAP} to derive this result. We note that
in Eq.~\eqref{eq:id-dagger} the sum over $j$ and $k$ may be read as a
trace operation, so that
\begin{align*}
       \mathsf{Id}(A)\ =\  & \frac{1}{d}\
                  \sum_{lm}\ \Tr\left(\gd_{lm}^{\dagger} A \right)
                               \gd_{lm}
\ \ ,
\end{align*}
that is, we recover Eq.~\eqref{eq:Blochdec}, the matrix decomposition 
of the operator $A$.
%
%
\subsection{General linear maps}
\label{sec:linear}
%
Extending the results of the preceding paragraphs we may
analyze the matrix expansion of general linear maps 
$\mathcal{L}(A)$, $A\in\mathcal{B}(\mathcal{H}_d)$ (also
referred to as superoperators). It is natural to study a linear map
via the action on a basis of the Hilbert-Schmidt space. Here, this means to
analyze, e.g., the Bloch representation of $A$,
\begin{align}
     \mathcal{L}(A)\ =\ & \mathcal{L}
              \left( \frac{1}{d}\sum_{jk} \gd_{jk}\Tr\left[\gd_{jk}^{\dagger}A\right]
                   \right)
\nonumber\\
                     =\ & \frac{1}{d}\ \sum_{jk}\ \Tr\left(\gd_{jk}^{\dagger}A\right)\ 
                                 \mathcal{L}\left(\gd_{jk}\right)
\nonumber\\
 =\ & \frac{1}{d^2}\ \sum_{jkab}\ \gd_{ab}^{\dagger}\cdot\gd_{jk}^{\dagger}\cdot A\cdot
                               \gd_{ab}\cdot \mathcal{L}\left(\gd_{jk}\right)\ \ ,
\end{align}
where we have used the expansion of the trace map in the last line. This
representation is interesting if the action of $\mathcal{L}(\cdot)$ 
on the matrix basis $\{\gd_{jk}\}$ is simple. For example, it is easy
to recognize the result for the identity map.

It is worthwhile mentioning here also the 
Choi-Jamiolkowski representation~\cite{Wolf2012,Schuch2016,Bengtsson2017} 
of the 
map $\mathcal{L}(\cdot)$, because it is obtained via a similar reasoning.
The Choi state $C_{\mathcal{L}}$ is defined as
\begin{align}
      C_{\mathcal{L}} =
      \mathcal{L}\otimes \mathsf{Id}\ket{\Phi^+_d}\!\bra{\Phi^+_d}
       = \frac{1}{d^2}\ \sum_{jk}\ \mathcal{L}\left(\gd_{jk}\right)\otimes
                                                    \gd_{jk}^{\ast}\ .
\end{align}
Then the map $\mathcal{L}(A)$ can be represented as
\begin{align}
 d\ \Tr_{[2]}\left[ C_{\mathcal{L}}\cdot\left( \id_d\otimes A^T\right)\right]&\ =\ 
         \frac{1}{d}\
         \sum_{jk}\ \mathcal{L}\left(\gd_{jk}\right)\ \Tr\left( \gd_{jk}^* A^T \right)
\nonumber\\
 =\ & \mathcal{L}\left( \frac{1}{d}\sum_{jk}\gd_{jk}\Tr\left[\gd_{jk}^{\dagger} A\right]
                      \right)
\nonumber\\
      =\ & \mathcal{L}(A)\ \ .
\end{align}
%
%
\subsection{The transposition}
\label{sec:transpose}
%
The examples in the previous paragraphs indicate that along these lines
it is possible to derive other results. The first is an explicit decomposition
of the transposition map. Taking into account the expansion of an operator $A$
in the standard basis, $A=\sum_{jk} a_{jk} \ket{j}\!\bra{k}$, 
we find~\cite{ES2018}
\begin{align*}
      A^T\ & =\ \sum_{jk} \ a_{kj} \ket{j}\!\bra{k}
\nonumber\\
           & =\ \sum_{jklm}\ a_{lm}     \ket{j}\!\bra{k}\cdot\ket{l}\!\bra{m}
                                                        \cdot\ket{j}\!\bra{k}\ \ .
\end{align*}
If we want to apply the transformation Eq.~\eqref{eq:Ustandard}, we have to
take one of the basis matrices in the decomposition with a complex conjugation,
\begin{align}
      A^T\ & =\ \frac{1}{d}\ \sum_{jklmpq}\  U^*_{jk,lm} \gd_{lm}\cdot A
                                                        \cdot \gd_{pq}^* U_{jk,pq}
\nonumber\\
           & =\ \frac{1}{d}\ \sum_{lm}\  \gd_{lm}\cdot A \cdot \gd_{lm}^*  \ \ .
\label{eq:transpose}
\end{align}
We notice that this identity provides the link between the four-operator
product in Eq.~\eqref{eq:fourops2} and Eq.~\eqref{eq:skewSWAP}.
%
%
%
\subsection{Partial transpose and  reshuffling operation}
%
Transposition has several generalizations if one considers
multi-party operators~\cite{Oxenrider1985,Karol2004,Karol2013,Bengtsson2017}. 
The first of them is partial transposition,
for example, on the second party. Consider the two-party operator
$B\in \mathcal{B}(\mathcal{H}_d\otimes\mathcal{H}_d)$, that is,
\begin{align*}
    B\ =\ \sum_{jklm}\ B_{jk,lm} \ket{jk}\!\bra{lm}\ \ .
\end{align*}
Then partial transposition on the second party exchanges 
the second indices in both groups,
\begin{align}
      B^{T_2}\ =\ \sum_{jklm}\ B_{jm,lk} \ket{jk}\!\bra{lm}\ \ ,
\end{align}
while the first party remains unaffected. By using the results
of the previous section we can give a matrix expansion for this map,
\begin{align}
       B^{T_2}\ =\ \frac{1}{d}\ \sum_{jk}\ \left(\id_d\otimes \gd_{jk}\right)
                                \cdot B \cdot 
                                           \left(\id_d\otimes \gd_{jk}^*\right)
  \ \ .
\label{eq:PTmap}
\end{align}
Alternatively, one can apply the reshuffling map  to $B$ that 
exchanges the second index in the first  with the first index
in the second group,
\begin{align}
      B^{R}\ =\ \sum_{jklm}\ B_{jl,km} \ket{jk}\!\bra{lm}\ \ .
\end{align}
In full analogy with what was said before we have
\begin{align}
       B^{R}\ =\ \frac{1}{d}\ \sum_{jk}\ \left( \id_d\otimes\gd_{jk}\right)
                                \cdot B \cdot 
                                           \left( \gd_{jk}^*\otimes\id_d\right)
  \ \ .
\label{eq:Rmap}
\end{align}
This is readily extended to the case of more than two parties, where 
any subset of indices in the first group can be exchanged with another
subset of the same size in the second group.
%
%
%
%
%
%
\subsection{The universal state inversion}
%
For completeness, we include the universal state inversion
map here, because it provides an example where a specific choice
of matrix basis leads to an interesting expansion. Universal
state inversion is a relevant map that applied to physical states
helps to explore the boundary of the state space and is therefore
also useful in entanglement detection~\cite{Horodecki1999,Rungta2001,Hall2005,ES2018,Eltschka2018}.

Consider first a Hermitian single-party operator $A$, i.e.,
$A\in \mathcal{B}(\mathcal{H}_d)$ and $A=A^{\dagger}$.
Universal state inversion is defined as
\begin{align}
   \mathcal{S}(A)\ =\ \Tr(A) \id_d\ -\ \mathsf{Id}(A)\ \ .
\end{align}
Because of the hermiticity of $A$ we can write $A=\left(A^*\right)^T$.
By using the results for the trace and the transposition maps we can
rewrite the definition
\begin{align}
   \mathcal{S}(A)\ & =\ \frac{1}{d}
                        \sum_{jk} \gd_{jk}\cdot A^*\cdot \gd_{jk}^{\dagger}
             \ -\ \frac{1}{d}\sum_{lm} \gd_{lm}\cdot A^* \cdot \gd_{lm}^*
\nonumber\\
                   & =\ \frac{1}{d}\ \sum_{jk}
                          \gd_{jk}\cdot A^* \cdot \left(
                               \gd_{jk}^{\dagger}-\gd_{jk}^*\right)\ \ .
\label{eq:matrix-state-inv}
\end{align}
For the Gell-Mann basis we have the special situation that the 
matrices are Hermitian and have either only real elements, or only
imaginary ones. That is, the paranthesis in Eq.~\eqref{eq:matrix-state-inv}
is non-zero only for the imaginary Gell-Mann matrices
$\Lambda^{(d)}_{lk}\equiv \yd_{kl}$, cf.~Eq.~\eqref{eq:GMY}.
Therefore
\begin{align}
   \mathcal{S}(A)\ 
                   & =\ \frac{2}{d}\ \sum_{j<k}
                          \yd_{jk}\cdot A^* \cdot 
                               \yd_{jk}\ \ .
\label{eq:Y-state-inv}
\end{align}

This result is readily generalized to multi-party operators.
In order to present the idea it suffices to consider the case of
two parties, i.e., Hermitian operators  
$B\in \mathcal{B}(\mathcal{H}_d\otimes\mathcal{H}_d)$. Then
\begin{align}
   \mathcal{S}(B) &  = \big[\Tr_{[1]}(\cdot)\otimes \id_d-\mathsf{Id}\big]
                   \cdot
                  \big[\Tr_{[2]}(\cdot)\otimes\id_d - \mathsf{Id}\big]\ B
\nonumber\\
       & =\ \Tr_{[12]}\left(B\right) \id_{d^2}
             - \Tr_{[2]}\left(B\right)\otimes\id_d\ -
\nonumber\\
            & \ \ \ \ \ \ \ \ \   \ \ \ \ \ \ \ \ \  \ \ \ \  
             - \id_d\otimes\Tr_{[1]}\left(B\right)
            + B\ \ .
\end{align}
By applying the same steps as in the single-party case we arrive at
\begin{align}
   \mathcal{S}(B) 
                   & = \frac{4}{d^2} \sum_{j<k,l<m}
                          \left(
                          \yd_{jk}\otimes\yd_{lm}\right)
                          \cdot B^* \cdot 
                               \left(
                               \yd_{jk}\otimes\yd_{lm}\right)\  .
\label{eq:Y-state-inv2}
\end{align}
We note that this is reminiscent of Wootters' $R$ matrix for two
qubits~\cite{Wootters1998}, in fact, it is the higher-dimensional 
generalization of that definition~\cite{Rungta2001,ES2018}. 
For pure states $B=\ket{\psi}\!\bra{\psi}$ it directly leads
to the squared concurrence of $\ket{\psi}$,
\begin{align}
     \mathcal{C}(\psi)^2\ =\ & \Tr\big[ \ket{\psi}\!\bra{\psi} 
                          \mathcal{S}\left(\ket{\psi}\!\bra{\psi}\right) 
                                \big] 
\nonumber\\
                         =\ & \frac{4}{d^2}\sum_{j<k,l<m}
            \left| \bra{\psi} \yd_{jk}\otimes\yd_{lm} \ket{\psi^*} \right|^2
\ \ .
\end{align}
%
%
%

%

\section{Conclusion}
%
We have presented a few important definitions regarding orthogonal
bases of the Hilbert-Schmidt space and then derived various identities, 
that is, matrix
expansions of relevant operators and maps. We hope this brief account
renders the known relations for matrix bases more transparent 
and makes the derivation of similar identities easier.
There are many additional connections between the central
equalities derived here, and therefore also alternative ways to obtain them. 
We have not mentioned all of them
in order to avoid excessive cross-referencing and to keep the
line of reasoning visible.
Finally, we believe that this collection of results may give some intuition
for the structure of  matrix expansions and for the  typical applications
where they have proven useful.
\vspace*{3mm}
\acknowledgments
Helpful comments on the manuscript by Christopher Eltschka and Marcus Huber are 
gratefully acknowledged.  C.\ Eltschka suggested to
include Sec.~\ref{sec:linear}. The author is grateful
to Karol \.Zyczkowski for various improvements in the manuscript and
for suggesting an appropriate title.
This work was supported
by grant PGC2018-101355-B-100 funded by 
MCIN/AEI/ 10.13039/501100011033 and by "ERDF A way of making Europe"
and Basque Government grant IT986-16.
%
%

%
%
%

\end{document}